\documentclass[twocolumn,prl,aps]{revtex4-1}

\usepackage{color,latexsym}
\usepackage{graphics}
\usepackage{epsfig}
\usepackage{graphicx,color,epsfig,rotate}

\begin{document}
\noindent
{\large \bf Comment on ``Two-spinon and four-spinon continuum in a frustrated
ferromagnetic spin-1/2 chain''} 

\vspace{0.1cm}

Recently Enderle {\it et al.} reported
 an inelastic neutron scattering (INS) study of the
dynamic spin susceptibility Im$\chi(\omega,k)$ for
LiCuVO$_4$ 
\cite{Enderle10}. Therein 
they
claim that 
(i) LiCuVO$_4$ 
is well described by
two interpenetrating, {\it weakly } ferromagnetically (FM)
coupled 
Heisenberg antiferromagnetic
spin-1/2 chains (HAF), (ii) the obtained exchange integrals
$J_i$ (NN and NNN inchain couplings $J_1$=$-$~1.6~meV,
$J_2$=3.56~meV, diagonal interchain coupling $J_5$=$-$0.4~meV in
the (ab)-plane, $\alpha$=$-J_2/J_1 \sim 2.2$) agree with those from
an analysis based on spin-wave theory (SWT) \cite{Enderle05},  (iii) the {\it observed} 
INS intensity above 10 meV belongs to a 4-spinon continuum (4SC).
Applying exact diagonalization (ED) and DMRG 
methods
to fit 
INS and magnetization $M(H)$
data, supported by 
independent microscopic
methods \cite{remarklast}, 
we will show that 
the 
claims of Ref.~\cite{Enderle10} are not justified
and that 
LiCuVO$_4$ 
exhibits 
$\alpha <$1, i.e. {\it strong} coupling of the HAF, 
at odds with (i).
For possible
spin nematics 
and Bose condensation of 2-magnon bound states 
in 
LiCuVO$_4$ 
\cite{Zhitomirsky10,Svistov10} precise knowledge of
the coupling regime is of 
key
importance.  

 Starting 
 from
 a 2D model,
the authors
suggest an effective 1D-model with 
$J_{\mbox{\tiny eff,}1}$=$J_1+2J_5$=$-$2.4~meV and   
a renormalized $J_{\mbox{\tiny eff,}2}$=$2J_2/\pi$, 
i.e.\
$\alpha$$\simeq$1.4.   
But the applied perturbative method is designed for $\alpha \gg$1.
Further problems
occur 
for the dispersion of 
spin excitations $\omega (k)$ (dark red curve in Fig.\ 2  of Ref.\ 1).
Near $k$=$\pi/4$ the local maximum  $\Omega$ gives 4.84~meV.
For  $\alpha$=1.42 and $J_{\mbox{\tiny eff,}1}$=$-$2.4~meV one has $\Omega$=4.36  (4.79)~meV, only,
according to our dynamical DMRG (ED) calculation with $L$=96 (28) 
sites of Im$\chi(\omega,q)$. 
Fitting our DMRG results for $0.5 \leq \alpha \leq 2$, 
the general
constraint for $\Omega(\alpha)$
 reads:
\vspace{-0.1cm}
\begin{equation}
\vspace{-0.1cm}
 4.84 \mbox{meV}=
 \Omega =J_{\mbox{\tiny eff,}1}\left(0.573-1.702\alpha +0.0109\alpha^2\right).
 \vspace{-0.05cm}
\end{equation}
More strikingly, 
fitting the
$M(H)$-data at
$T$=1.6~K \cite{Enderle05,Drechsler07}
by our  DMRG ($T$=0, $L$=512~sites),
we found
$\alpha$ =0.75 
(see Fig.\ 1a). 
We obtain 
$H_{\mbox{\tiny s}}$=41.6~T, $g$=2.27 and in accord with Ref.\ 5 
 d$\frac{M}{M_{\mbox{\tiny s}}}$/d$\frac{H}{H_{\mbox{\tiny s}}}$$\approx$0.39 
 at low $H$,  where  $H_{\mbox{\tiny s}}(M_{\mbox{\tiny s}})$
 is the saturation field (magnetization). Then, Eq.\ (1)
 gives  $J_{\mbox{\tiny eff,}1}$=$-$6.95~meV and $J_{\mbox{\tiny eff,}2}$=5.2~meV.  
 The 1D-set of Ref.\ 1 yields a too small $H_{\mbox{\tiny s}}$=37.3~T at $g$=2.27
 or a too small $g$=2.03 for $H_{\mbox{\tiny s}}$=41.6~T, only, and 
 clearly
 too high $M$
 above 0.6$H_{\mbox{\tiny s}}$ (see Fig.\ 1a). 
If the SWT-fit is 
meaningful,
 $J_1$
 is strongly renormalized but \ $J_{\mbox{\tiny eff,}2}$ is close to its
  bare value of 5.6~meV \cite{Enderle05} both
  at odds with
  Refs.\ 1,2.
  
  With our
  fitted
  $J_{\mbox{\tiny eff,}2}$-value, almost the whole region ascribed 
  to the 4SC \cite{Enderle10} is covered now by the extended 2-spinon continuum (2SC).
  The 4SC should be looked for at $k=0.5$ above 16.3~meV (see Fig.\ 1c) 
  i.e.\ in a region which has 
  {\it not} been measured yet.
 %
 Since most of the 
 INS intensity below 16~meV
  belongs to the 2SC, 
  the 
  size of the 4SC
  enhancement
  compared to that of a HAF is not yet settled
  and the 4SC 
  in Fig.\ 4 of Ref.\ 1 is overestimated.
  Figs.\ 1b, 1c show
that the dispersion of the 
INS peaks
is insufficient to find a unique
$J_1$-$J_2$ set. 
INS intensities
or other quantities like $M(H)$
must be analyzed, too. Our set 
explains the larger 
 INS intensity above 9.5~meV as compared with
 that
in Ref.\ 1 (see the boxes in Fig.\ 1).

To conclude, a weakly coupled HAF model with small 
$|J_1| \leq$ 5~meV
 as in Ref.\ 1 
is not justified 
for 
LiCuVO$_4$
whereas a strong coupling regime 
with $|J_1|$$ >$$ J_2$ where both $J$'s exceed significantly those 
of Ref.\ 1
is consistent with  the INS, the magnetization, 
and realistic 
 microscopic models  of edge-shared cuprates \cite{remarklast}.
For strongly coupled HAF's, 
the 2SC is extended to higher $\omega$ with
less spectral weight left for the 4SC.
Future 
studies are highly desired
to refine the main $J$'s and the 4SC as well.
\begin{figure}[b]
\includegraphics[width=7.8cm]{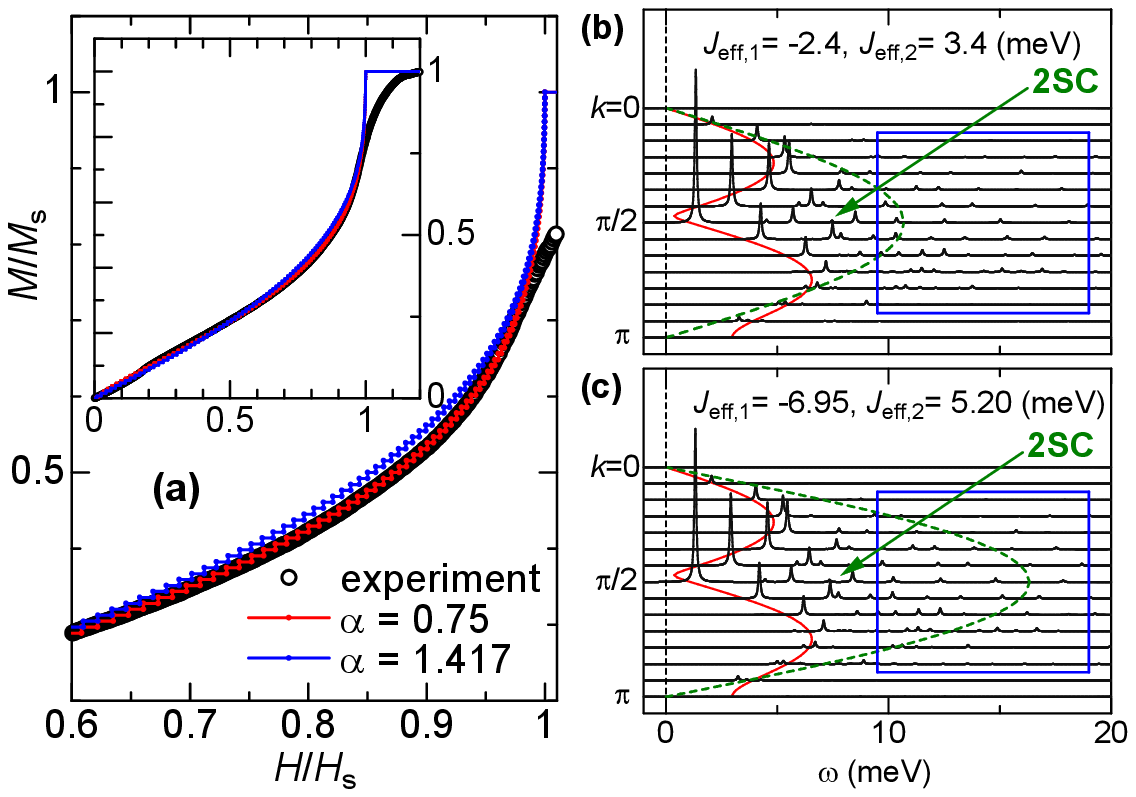}
\caption{Magnetization of LiCuVO$_4$ 
for the 1D $J$-set
of Ref.\ 1 and our fit compared with experiment ($H \parallel c$ ) (a).
Im$\chi(\omega, k)$ from exact diagonalizations for both sets
for a chain with $L$=28 sites broadened with 0.05~meV. Red curves: dispersion
of spin excitations from main INS peaks \cite{Enderle05}(b,c).
Notice a slight downshift near $k$=$\pi$/4 
of 0.426 meV and 0.676 meV for Fig.\ 1b(1c),
respectively, within the DMRG (see Eq.\ (1)).}
\label{chi}
\end{figure}
\vspace{0.1cm}

\noindent
{\small 
S.-L. Drechsler$^1$, S.\ Nishimoto$^1$, R.\ Kuzian$^1$, J.\ M\'alek$^{1,2}$,
W.\ Lorenz$^1$, 
J.\ Richter$^3$, J.\ v.\ d.\ Brink$^1$, M.\ Schmitt$^4$, 
H.\ Rosner$^4$}\\ 
\indent 
{\small $^1$ Inst.\  f.\ Festk\"orper- u.\ Werkstoffforschung IFW-Dresden}\\
\indent
\small{01171 Dresden, Germany, email: drechsler@ifw-dresden.de}\\
\indent
\small{ $^2$ Institute of Physics, ASCR, Prague, Czech Republic}\\
\indent
\small{ $^3$ Inst.\ f.\ Theor.\ Physik, Universit\"at Magdeburg, Germany.}\\
\indent
\small{ $^4$ MPI-CPfS Dresden, Germany.}

\vspace{0.25cm}
\noindent
\noindent
\noindent


\begin{references}
\bibitem[3]{remarklast}A strong coupling regime with $\alpha $$<$1 is 
also
supported by
independent microscopic calculations that predict 
$J_1$=$-$6.3~meV, $J_2$= 5.1~meV, $\alpha$= 0.8 (mapping a 5-band
Hubbard model) and $J_1$=$-$8.8~meV, $J_2$=
6.5~meV, $\alpha$= 0.74 (LSDA+$U$). Both methods 
as our phenomenological analysis given above
are
valid irrespective 
of the value of $\alpha$ 
and have been applied succesfully
for many related systems.
Note that in all 
systems 
with a
 Cu-O-Cu bond angle 
 {\tiny $ \stackrel{ <}{\sim}$}95$^\circ$ we found a
FM NN coupling and $|J_1|$
$\gg$ 1.6~meV 
\cite{Drechsler10}.
\end{references}
\end{document}